# A Hybrid Scenario for Gas Giant Planet Formation in Rings

Richard H. Durisen

Department of Astronomy, Indiana University, 727 E. 3rd St., Bloomington, IN 47405-7104

E-mail: durisen@astro.indiana.edu

Kai Cai

Department of Astronomy, Indiana University, 727 E. 3rd St., Bloomington, IN 47405-7104

Annie C. Mejía

Department of Astronomy, University of Washington, Box 351580, Seattle, WA 98195-1580

Megan K. Pickett

Department of Chemistry and Physics, Purdue University Calumet, 2200 169th St., Hammond, IN 46323-2094

**Proposed Running Title:** Gas Giant Planet Formation in Rings

**Editorial correspondence to:**

Dr. Richard H. Durisen
Department of Astronomy
Indiana University
727 East 3rd Street
Bloomington, IN 47405-7104
Phone: 812-855-6921
Fax: 812-855-8725
E-mail: durisen@astro.indiana.edu

## ABSTRACT

The core-accretion mechanism for gas giant formation may be too slow to create all observed gas giant planets during reasonable gas disk lifetimes, but it has yet to be firmly established that the disk instability model can produce permanent bound gaseous protoplanets under realistic conditions. Based on our recent simulations of gravitational instabilities in disks around young stars, we suggest that, even if instabilities due to disk self-gravity do not produce gaseous protoplanets directly, they may create persistent dense rings that are conducive to accelerated growth of gas giants through core accretion. The rings occur at and near the boundary between stable and unstable regions of the disk and appear to be produced by resonances with discrete spiral modes on the unstable side.

# I. INTRODUCTION

The theory of gas giant planet formation is at a critical juncture between the two major theories – core accretion (Perri and Cameron 1974, Mizuno 1980) and disk instability (Kuiper 1951, Cameron 1978). In its classic form, core accretion works sufficiently fast for Jupiter and Saturn in our own Solar System but encounters time scale problems for Uranus and Neptune (Pollack et al. 1996, Wuchterl et al. 2000). As argued by Boss (e.g., 2001), the relatively short disk lifetimes in star forming environments (e.g., Briceño et al. 2001, Haisch et al. 2001, Lada and Lada 2003) and the nature of observed extrasolar planets, which include hot Jupiters close to their stars as well as super-Jupiters of many Jupiter masses ($M_J$) at greater distances (e.g., Marcy et al. 2003, Udry et al. 2003), pose difficult challenges for the core-accretion model. This situation has been made even more difficult recently by indirect evidence for a planet in the $10^6$ year old star/disk system CoKu Tau 4 (Forrest et al. 2004). Migration in a laminar nebula may prevent gas giants from achieving multiple Jupiter masses by core accretion (Terquem et al. 2000, Nelson et al. 2000). However, the magnitude and even direction of migration is now subject to doubt (Masset and Papaloizou 2003, Nelson and Papaloizou 2003, 2004, Terquem 2003). Gas giants can be formed in close orbits by core accretion without migration (Bodenheimer et al. 2000, Ikoma et al. 2001), and time scales for core accretion can be shortened if dust opacities in the accreting gas envelopes are lowered (Lissauer 2001, Hubickyj et al. 2003) and if migration becomes a random walk due to turbulence (Rice and Armitage 2003, Laughlin et al. 2004).

As reviewed in Durisen (2001) and Durisen et al. (2003), modern 3D hydrodynamics simulations have shown that gravitational instabilities (GI's) in protoplanetary disks can lead to fragmentation of the disk into dense spiral arcs and clumps for sufficiently short cooling times (Boss 2001, 2002, 2003, 2004, Gammie 2001, Rice et al. 2003a, 2003b, Johnson and Gammie 2003, Mejía et al. 2004, hereafter MDPC) or under isothermal conditions (Nelson et al. 1998, Pickett et al. 1998, 2000a, 2000b, Boss 2000, Johnson and Gammie 2003, Mayer et al. 2002, 2003, 2004). Boss (1997, 1998) was the first to revive the idea that such clumps could become permanently bound and evolve into gas giant planets. For isothermal disks, a few global 3D hydrodynamics calculations have been followed long enough to show dense multi-Jupiter mass clumps surviving for many (Boss 2000) or even very many (Mayer et al. 2002, 2004) orbits. The beauty of the disk instability mechanism is that, if it works, it can produce multi-Jupiter mass planets and even entire planetary systems (Mayer et al. 2002, 2004) within a few tens of disk orbits, obviating disk lifetime problems. Nevertheless, it remains to be demonstrated that clumps will survive as bound gaseous protoplanets in disks with realistic thermal physics. Boss (2001, 2002) has included cooling by radiative diffusion, but none of his 3D hydro disk simulations are integrated for long enough times at high enough resolution to be conclusive about clump survival. Other researchers (Nelson et al. 2000, Mejía et al. 2003, Mejía 2004, Ph.D. Dissertation) generally do not see dense clumps when using alternative treatments of radiative cooling. In the best of our own simulations published to date (Pickett et al. 2003, hereafter PMD) (the isothermal case with 256 azimuthal zones), we do not see permanent bound protoplanetary clumps. We admit that 256 azimuthal zones may be insufficient to allow clumps to survive in a grid-based calculation (Boss 2000), but clumps are also not permanent in our more recent simulations (MDPC) with 512 azimuthal zones. The extreme clump longevity in smoothed particle hydrodynamics (SPH) simulations (Mayer et al. 2004) needs to be confirmed in grid-based codes to ensure that it is not an artifact of numerical features unique to SPH. Although the work of Boss and Mayer et al. is suggestive, we feel it is premature to conclude that disk fragmentation leads to protoplanets.

The purpose of this short paper is to focus on a particular feature of the PMD and MDPC simulations in light of the ongoing debate about gas giant planet formation. We suggest that, even if GI's do not lead to permanent clump formation, they may significantly accelerate core accretion by creating persistent dense gas rings near boundaries between GI active and inactive regions. As an interesting hybrid of the core-accretion and disk instability planet formation theories, we feel it warrants being brought to the attention of the planetary science community quickly in a separate paper.

## II. SIMULATIONS

PMD discuss two relevant grid-based 3D hydrodynamics simulations of disks around young stars. As explained in that paper, it is possible to scale our disk models to different masses and radii. For this paper, the initial disk will be assumed to extend from 3 to 40 AU and orbit a star with a mass $M_s = 1 M_\odot$. The disk mass $M_d$ is $0.14 M_\odot$ and is distributed so that initially the surface mass density $\Sigma \sim r^{-1/2}$ except near the edges. With such a large mass, the star/disk should probably be construed as being very young, possibly still in its embedded phase. Let $\varepsilon$ be the internal energy density and $\Lambda$ the volumetric cooling rate. Cooling is characterized by a time scale $t_{cool} = \varepsilon/\Lambda$ which is chosen to be constant and equal to 2 orps everywhere that it is applied. One "orp" or "outer rotation period" is 179 yrs and corresponds to the orbit period at about 33 AU. In the PMD calculation called HighQ-HC-Full, here referred to simply as "Full", the initial disk is marginally stable to GI's with a minimum Toomre Q of 1.8, and the cooling is turned on throughout the disk during the simulation. For the LowQ-HC-Half case in PMD, which we will call "Half", the initial outer disk is strongly GI unstable (Q ~ 1), and cooling occurs only in the outer half of the disk. The only source of dissipative internal heating for both simulations is artificial bulk viscosity in regions of strong compressions, such as shocks. The initial disk is quiescent and in hydrodynamic equilibrium with no heat sources. Due to cooling, the Full disk cools to instability in its outer region and develops a strong four-armed spiral by 4 orps (717 yrs). The instability appears sooner in the Half disk. In PMD, the Half case is followed for 10.4 orps (1,864 yrs), and the Full case for 16.6 orps (2,975 yrs).

By design, Half has a hot, high-Q, stable inner disk with little nonaxisymmetric structure. By the end of the simulation, the outer disk settles into a nearly steady-state level of GI activity that drives mass inward. Because the inner disk cannot sustain transport, a persistent dense ring with strong but transient dense clumps forms at the boundary (r ≈ 20 AU) between the GI-active and inactive regions (see Figure 7b of PMD).

By about 12 orps, the Full simulation in PMD establishes a dynamic quasi-equilibrium of GI activity where shock heating by spiral waves balances the $t_{cool} = 2$ orp cooling everywhere outside r ≈ 10 AU to maintain an overall average Q ≈ 1.45. To verify that this is a quasi-steady state, MDPC have now continued this integration out to 23.5 orps (4,211 yrs) and find that this asymptotic behavior persists. New insights have been gained by analyzing this lengthened simulation. Fourier decomposition of the complex nonaxisymmetric spiral structure in the r > 10 AU region, as discussed in PMD and MDPC, shows that this GI activity is dominated by a few discrete underlying m = 2 (two-armed) patterns with Corotation Radii (CR's) near 26 and 29 AU. Averaged over long times, there is a net steady mass inflow inside 29 AU at a rate of about $10^{-6}$ $M_\odot$/yr due in great part to these modes. Although the global two-armed modes appear to dominate transport, there is significant power in all azimuthal Fourier m-values with a complex pattern of periodicities indicative of extremely strong nonlinear coupling, referred to as "gravito-turbulence" by Gammie (2001), which involves dozens to hundreds of modes for each m over the m-range analyzed (m = 1 to 6) (see also Laughlin et al. 1997, 1998). In MDPC, the dependence on the cooling time $t_{cool}$ for various characteristics of the asymptotic state are tested by rerunning the simulation from 11 to 18 orps with $t_{cool}$ decreased from 2 orp to 1 orp. The Fourier mode amplitudes of the quasi-steady nonaxisymmetric structure and the inward mass transfer rate for r < 29 AU increase by corresponding factors.

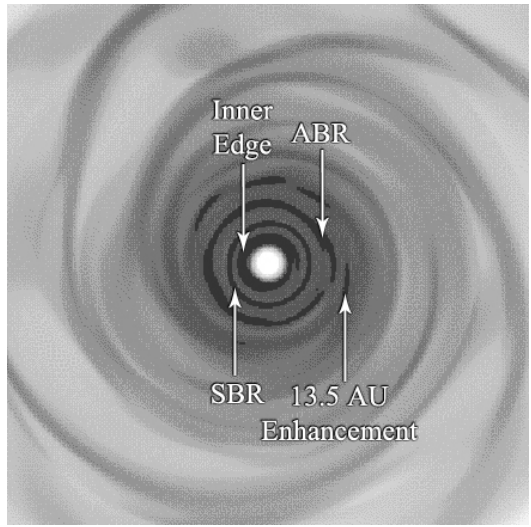

**Figure 1.** Logarithmic grey scale representation of the midplane density for the inner disk of the $t_{cool}$ = 2 orp Full simulation at 22.0 orps (3,942 yrs). Each side of the panel is 84 AU in length. Arrows indicate several features discussed in the text. SBR stands for "secondary boundary ring", and ABR for "active boundary ring".

The inner disk of Full also reaches a quasi-equilibrium balance of heating and cooling but maintains a GI-stable value of Q. Except for localized edge modes at the inner edge and important resonances with r > 10 AU patterns, the inner disk region is not GI-active but is dominated by three dense, nearly axisymmetric rings (see Figures 1 and 2). The density peak at 3 AU is just the inner edge itself. The ring at about 9.5 AU sits at the boundary between the outer GI-active low-Q region and the inner GI-inactive region and is similar dynamically to the ring in the Half calculation. Although on average this is a ring-like surface density enhancement, it contains a semi-persistent three or four-clump structure, where the individual clumps come and go over several orbits. These clumps do not have extremely high contrast with their surroundings like the isolated clumps and arcs found in a fragmented disk. The middle peak at r ≈ 6.7 AU remains nearly, but not quite, axisymmetric and grows steadily in peak density and mass from about 12 orps onward, as shown in Figure 2. By 23.5 orps, it is about 1.5 AU in full width, contains about 6 $M_J$, and is still growing. The ring peaking at r ≈ 9.5 AU is 3 AU in full width, contains 18 $M_J$, and maintains a quasi-steady structure. When $t_{cool}$ is decreased from 2 orps to 1 orp, the 6.7 AU ring grows about twice as fast, reaching the same mass by 18 orps that the $t_{cool}$ = 2 orp disk reaches by 23.5 orps.

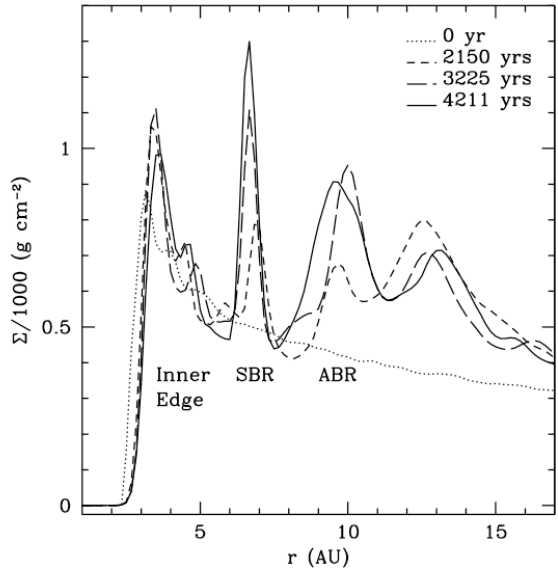

**Figure 2.** Plot of the azimuthally averaged disk surface density $\Sigma(r)$ that illustrates the growth of the rings in the $t_{cool}$ = 2 orp FULL simulation. The times shown correspond to t = 0, 12, 18, and 23.5 orps.

## III. POSSIBLE CAUSES FOR THE RINGS

We will refer to the outermost of the three rings (r ≈ 9.5 AU) in the Full calculation as an "active boundary ring" (ABR), because it occurs at the boundary between GI-active and inactive regions and because it displays active nonaxisymmetric dynamics, namely, the semi-persistent dense clumps. It is well known that gravitational torques due to GI's produce inward transport of mass (Larson 1984, Boss 1984, Durisen et al. 1986). Since there is no global mechanism in the inner hot regions to continue the transport inward, mass piles up at the boundary on the GI-active side with an ongoing multi-clump nonaxisymmetry. Disks are expected to have Q-distributions that decrease outward (e.g., D'Alessio et al. 1999), and so disks experiencing GI's should have boundaries between GI-active and inactive regions. This could also happen at a few AU if matter is piling up in a dead zone (Gammie 1996, Armitage et al. 2001). How strong the enhancement of density will be in the ABR probably depends on the efficacy of other transport mechanisms in the GI-inactive region. No such mechanisms have yet been included in our simulations, and their effect on the ABR will be an interesting subject for future study. The ABR itself is well resolved (about 15 radial cells in Full), and we are confident it is a real physical phenomenon.

The middle ring (r ≈ 6.7 AU) in the Full simulation does not exhibit marked nonaxisymmetric structure, although it does occasionally contain one semi-persistent clump. For lack of better terminology, we will simply call it a "Secondary Boundary Ring" or SBR. We are not certain of its cause, but there may be contributions from two interrelated mechanisms – resonances with nonaxisymmetric structure and axisymmetric wave dissipation. As shown in Figure 3, Fourier analysis of time series data for density structure in the azimuthal direction over the time interval 18 to 23.5 orps reveals that the region between the 6.7 AU and 9.5 AU rings contains strong Inner Lindblad Resonances (ILR) for several persistent 2, 3, 4, and 5-armed patterns with CR's outside or in the ABR. Nonaxisymmetric waves generated at these ILR's will drain angular momentum from material in the region and cause it to pile up in the SBR that lies interior to these ILR's. Similarly, as illustrated in Figure 3, the inner part of the GI-active region seems to be sculpted by ILR's of discrete waves. The ILR's of several m = 2 modes with CR's beyond 20 AU lie outside and probably cause the radial concentration of mass at about 13.5 AU. This is not an axisymmetric ring but consists of several strong spiral features with CR near the same location (see Figure 1). The ILR for an m = 2 wave with CR at 14 AU lies just outside the SBR. The m = 3, 4, and 5 ILR's due to structure in the outer part of the ABR lie outside the SBR, and the m = 2 ILR for this structure, plus the m = 3 and 5 ILR's for structure in the ABR may be spawning another ring just inside the SBR. An analysis using time series data from 11 to 18 orps shows a similar but not identical pattern of ILR's. The ILR's associated with the strongest radial concentrations of mass do tend to be present in both time intervals.

The coincidence of strong resonances with radial concentrations in the azimuthally averaged $\Sigma(r)$, as shown in Figure 3, is a strong argument for the role of resonances in ring formation. Given the multiple resonances inside and outside the SBR, substructure might appear in this region if the calculation had higher resolution. However, the SBR itself is straddled by low-order resonances with strong signals and would probably remain a single strong ring. Complex interactions of discrete low-order modes acting over distances greater than the disk scale height (< 1 AU at these radii) mean that a local thin-disk treatment of GI's (Gammie 2001, Johnson and Gammie 2003) would miss essential features of GI activity in this disk (see also Laughlin and Rozyczka 1996, Balbus and Papaloizou 1999, MDPC).

Additional analysis suggests there may also be axisymmetric waves, with typical radial wavelength similar to the spacing between the rings, which penetrate into the inner disk through the ABR. They extend down to the SBR, but end there. It is clear that the strong nonlinear coupling in the GI-active region will generate axisymmetric waves (Laughlin et al. 1997, 1998). Both axisymmetric and nonaxisymmetric waves are probably best viewed as part of one nonlinear process. The Half simulation indirectly corroborates this interpretation. In Half, both artificial bulk viscosity and cooling are turned off inside the ABR, and no SBR appears. Although heating by dissipation of waves contributes to the high-Q thermal balance achieved in the inner disk region of Full, the values of Q in and near the SBR, though too high for strong nonaxisymmetric instability, are low enough to allow even nonaxisymmetric waves to penetrate from the GI-active side (see Figure 3).

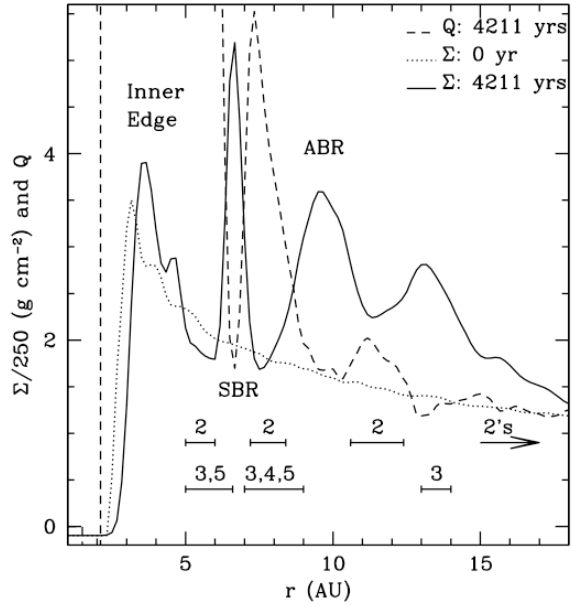

**Figure 3.** Comparison of Q(r) and azimuthally averaged surface density $\Sigma(r)$ with Inner Lindblad Resonance (ILR) locations. Here, the Toomre Q is $c_s\kappa/\pi G\Sigma$, where the sound speed $c_s$ and the epicyclic frequency $\kappa$ ($\approx$ the rotational angular frequency $\Omega$) are evaluated using azimuthally averaged quantities in the midplane. The horizontal lines show approximate locations of ILR's for strong patterns identified by Fourier analysis of the density in the azimuthal direction. The Corotation Radii (CR's) for the patterns are as follows: For the two-armed patterns, designated by a "2", the CR's are at approximately 11, 14, 16-19, and > 20 AU in order of increasing r. The horizontal band marked "3, 5" corresponds to the approximate ILR locations for three and five-armed modes with CR at 9.5 AU; the horizontal band marked "3, 4, 5" for three, four, and five-armed modes with CR at 10.5-12. The ILR marked "3" corresponds to a three-armed mode with CR at about 18.5 AU. Only the most distinct patterns are noted here. Power exists at other pattern periods, and the spectrum of modes becomes increasingly complex as both r increases and the number of arms increases.

The SBR in the Full simulation is only about 7 radial zones wide, and so we need to be cautious. It is possible that the ring itself or at least its detailed characteristics are artifacts of our 3D hydro code's von Neumann and Richtmyer artificial bulk viscosity scheme, the presence of an inner edge, or some other yet unidentified numerical problem. Our simulations are computer-intensive, and we have not performed extensive experiments to probe the dependence of the SBR on numerical parameters. However, we have rerun short stretches of the $t_{cool}$ = 1 orp simulation starting at 11.5 orps under different conditions. When we turn off both heating and cooling inside 10 AU, the SBR does not grow at all, as we expect on the basis of the Half calculation, which shows no SBR. Dissipation of disk energy is clearly necessary for ring growth. When we turn off only the heating by artificial bulk viscosity inside 10 AU, the SBR grows at a similar or perhaps somewhat faster rate. One might be tempted to conclude that wave dissipation is not involved in SBR growth. However, all finite difference codes have some dissipation introduced by truncation error terms. Also, as seen in Figure 3, Q in the ring itself is actually fairly low, and so self-gravity is important enough to enhance ring growth in the face of continued cooling once the ring is well established. We recognize the need for much more experimentation, especially at higher spatial resolution, and we plan to do so.

We have recently computed disk evolutions with realistic radiative cooling (Mejía 2004, Ph.D. Dissertation). The two longest simulations, for $M_s$ = 0.5$M_\odot$ and $M_d$ = 0.07$M_\odot$, one with stellar irradiation at the disk surface and one without, have so far only been computed to about 11 orps; but, in the case without stellar irradiation, a distinct ABR and an SBR are already evident at 10.6 orps with masses of 4.4 and 7 $M_J$, respectively, and they occur at similar disk radii as they do in Full. Overall disk cooling times are on the order of a few to tens of orps in the inner disk region, for the opacities used (D'Alessio et al. 1999). These opacities are similar to those in Boss (2001, 2002, 2004), but we do not see the fast cooling by convection and disk fragmentation that he reports. We currently suspect the discrepancy is due in part to different treatments of boundary conditions, and we are actively investigating this issue.

We will not discuss the inner edge ring (r ≈ 3 AU) further, because it is clearly an artifact of imposing a large inner disk radius on our initial disk model. Little mass inflow appears to occur interior to the middle ring, and the inner ring does not grow.

## IV. CONSEQUENCES FOR PLANET FORMATION

One simple effect of ring formation is the enhancement of surface density. In Full, at 23.5 orps (4,211 yrs), Σ at 6.7 AU and at 9.5 AU are 1,300 and 900 g cm$^{-2}$, which are enhancements by factors of 2.6 and 2.3, respectively, over the initial values at these radii. If the planetesimal surface density passively follows these enhancements, this would be sufficient to speed up gas giant formation by comparable factors (Greenwig and Lissauer 1992, Lissauer 1993). So gravity-induced enhancements over initial conditions that are ripe for GI's may also produce dense rings conducive to rapid core accretion.

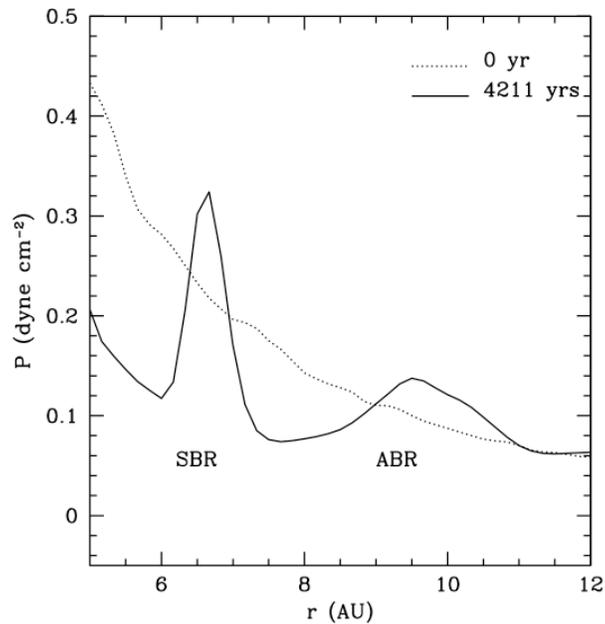

**Figure 4.** Plot of the azimuthally averaged pressure distribution P(r) in the midplane for the $t_{cool}$ = 2 orp Full simulation over the SBR and ABR region for t = 0 and 23.5 orps.

However, it is not density enhancement alone that may produce faster planet formation in rings. As shown in Figure 4, there are significant pressure maxima associated with the ring-like structures. As a result, any large solids produced before or during the GI-active phase will be drawn into the centers of the rings by drag forces. Thinking along similar lines, in light of Boss (2001, 2002, 2003) GI simulations, Haghighipour and Boss (2003, 2004, hereafter HB3a and HB3b) have begun to consider how solids will drift and accumulate against the background of high-contrast disk structure. It has been known for some time that solid particles orbiting in a gaseous disk drift radially in the direction of a radial pressure gradient (Weidenschilling 1977). Drift is fastest for particles of order meters in size. Using the simple "perturbed Keplerian flow" approximation discussed in Weidenschilling (1977) and using eqs. (4) and (8) of HB3a for the drag force, we calculate that particles with radii of 10 m near the foot of the SBR ring will drift radially by 1 AU, enough to drift into its peak, in only about 300 yrs. A similar analysis for the ABR gives 1 AU radial drift times of about $10^3$ yrs for the same particle size. Particles of order one meter in size should drift a factor of several 10's of times faster (see Figures 2 and 3 in HB3a). Note that these drift times are much shorter than the duration of our simulations! HB3b shows that vertical settling of meter-sized particles proceeds on similarly short time scales, and Haghighipour (2004) finds that growth of small particles occurs at an enhanced rate in the presence of a radial density enhancement. With 6 and 18 $M_J$ of gas within the SBR and ABR, respectively, up to 10 and 30 Earth masses ($M_E$) of solids could be concentrated into narrow regions near the ring pressure maxima.

This will have at least two effects favorable to core accretion – runaway growth times for solid planetary embryos will decrease by factors similar to the surface density enhancements of solids, and, as solids grow to meter sizes anywhere in the rings, they will rapidly drift into the feeding zone of the largest particle growing near the peak of the ring. It is at least plausible to imagine that a large fraction of solids in the rings can accumulate into a single solid core on a time scale much shorter than the several million years typically cited for core accretion. Similar acceleration of planet formation by disk structure has been suggested in other contexts, e.g., in vortices (Klahr and Henning 1997, Klahr 2003 private communication) and at edges of gaps (Bryden *et al*. 2000).

A potential problem in our case might be fluctuating gas velocities, especially in the ABR. Our simulations show that fluctuations from the mean flow are typically less than or on the order of the solid-gas relative velocities but do exceed the radial drift velocities of 10 m solids. More detailed calculations would be needed to determine the effect that random buffeting due to gas velocity fluctuations has on the concentration of solids. As the solids grow and deplete, radiative opacities should decrease substantially, and this will additionally enhance the rate of gas giant growth by core accretion (Hubickyj et al. 2003). The growth rate of the SBR itself will increase as $1/t_{cool}$ as dust growth decreases opacity, and Q in the ring will drop as the ring grows, which may eventually lead to GI-induced ring fragmentation. With all these processes working together, gas giant planet formation could become a runaway process.

Recent work on migration and gas giant formation in a turbulent disk (Nelson and Papaloizou 2003b, Rice and Armitage 2003, Laughlin et al. 2004) will need to be extended to GI active disks and to rings. Although the self-gravitating waves in the active region are constantly and perhaps chaotically changing, the dominant modes present are global in scale and may therefore produce systematic effects. The reduction of cooling time by growth of solids may make the entire disk susceptible to fragmentation (Gammie 2001, Johnson and Gammie 2003) in addition to accelerating ring growth at boundaries. Calculations combining gas dynamics with the growth, drift, and diffusion of solids are going to become necessary to understand planet formation by GI's.

## IV. CONCLUSION

On the basis of the evidence presented here from our numerical disk experiments, we propose a hybrid scenario for gas giant planet formation. Even if the occurrence of GI's does not lead directly to the formation of dense, bound protoplanets, density-enhanced rings may form near boundaries between GI-active and inactive regions due to resonances with discrete spiral modes in the active region combined with energy dissipation. Surface densities can increase by factors of several in just a few thousand years. Core accretion may then be accelerated in the rings by several effects working in concert – the enhanced gas densities themselves, the radial drift and concentration of meter-sized solids toward the ring peaks on time scales of only hundreds to thousands of years, the decrease in gas opacity due to growth and depletion of solids, and decreasing Q in the rings. It is unclear how formation of a core at the center of a dense narrow gas ring will affect migration. If migration is inhibited, then the rate of gas giant planet growth should depend primarily on how fast the core can accrete its gaseous envelope (Wuchterl et al. 2000) under these conditions.

We realize that there are a number of unresolved issues, so we consider this so far to be only a provocative suggestion. Particularly important tasks, which we hope to address in the near future, are to: 1) elucidate the ring formation mechanism, 2) test its dependence on numerical effects, and 3) study boundary regions when other mass transport mechanisms operate in the GI-inactive region. The growth of solids, their drift and accumulation, the growth of a gas giant by core accretion in the environment of either an ABR or a SBR, and migration in the presence of a dense, narrow, and possibly turbulent gas ring also need to be studied more carefully. Finally, it will be important to verify that ring formation can and will occur under realistic conditions. Our new calculations with radiative cooling are a start, but they must still be incorporated into a reasonable evolutionary scheme, which includes effects of the star/disk system environment, like irradiation, when appropriate. We have such efforts underway.

## ACKNOWLEDGMENTS

R.H.D. would like to thank his son Michael V. M. Durisen for assistance with the analysis of wave activity in the Full simulation. The authors also thank F.C. Adams, C. Gammie, referee A.P. Boss, and another anonymous referee for useful discussions and comments. R.H.D., K.C., and A.C.M. were supported in this research by NASA Origins Grant No. NAG5-11964. M.K.P. was supported by NASA Planetary Geology and Geophysics Grant No. NAG5-10262 and a Traveler's Grant from the University of Leeds. This work was also supported in part by systems obtained by Indiana University through its relationship with Sun Microsystems Inc. as a Sun Center of Excellence.